%% file: ms.tex
\documentclass[10pt,conference]{IEEEtran} 
\usepackage{graphicx}
\usepackage{amsmath}
\usepackage{algorithmic}
\usepackage{array}
\usepackage{url}
\usepackage{multirow}
\usepackage{subcaption}
\usepackage{makecell}
\usepackage{comment}
\usepackage{fancybox}
\usepackage{balance}
\graphicspath{{images/}}

\begin{document}
\title{Topics of Concern: Identifying User Issues in Reviews of IoT Apps and Devices}
\author{
   \IEEEauthorblockN{Andrew Truelove, Farah Naz Chowdhury, Omprakash Gnawali, and Mohammad Amin Alipour}
    \IEEEauthorblockA{Department of Computer Science, University of Houston, TX, United States}}

\maketitle
\input{texstuff}
\input{abstract}
\input{intro}
\input{related}
\input{method}

\input{issues}
\input{discussion}

\input{threats}
\input{conclusion}
\input{Acknowledgment}

\bibliographystyle{acm}
\bibliography{bib}
\appendix
\subsection{Results of topic modeling for IoT apps and devices}
\label{appendix}

\newpage

\input{extra}

\end{document}

%% file: texstuff.tex
\newcommand{\Fix}[1]{\textbf{Fix:#1}}
\newcommand{\Amin}[1]{\Fix{Amin:#1}}
\newcommand{\Part}[1]{\noindent\textbf{#1} }
\newcommand{\Comment}[1]{}

\newcounter{observation}
\newcommand{\observation}[1]{\refstepcounter{observation}
	\begin{center}
	\Ovalbox{
	\begin{minipage}{0.93\columnwidth}
		\textbf{Observation \arabic{observation}:} #1
	\end{minipage}
	}
	\end{center}
	\vspace{-5pt}
}

%% file: abstract.tex
\begin{abstract}
  Internet of Things (IoT) systems are bundles of networked sensors and actuators that are deployed in an environment and act upon the sensory data that they receive. 
 These systems, especially consumer electronics, have two main cooperating components: a device and a mobile app.
 The unique combination of hardware and software in IoT systems presents challenges that are lesser known to mainstream software developers.They might require innovative solutions to support the development and integration of such systems.   

  In this paper, we analyze more than 90,000 reviews of ten IoT devices and their corresponding apps and extract the issues that users encountered while using these systems.
  Our results indicate that issues with \emph{connectivity}, \emph{timing}, and \emph{updates} are particularly prevalent in the reviews.
  Our results call for a new software-hardware development framework to assist the development of reliable IoT systems.

\end{abstract}

%% file: intro.tex
\section{Introduction}
Internet of Things systems (IoT) are sets of interconnected sensors and actuators that are potentially backed and managed by servers on the Internet.
These systems are becoming part of ``smart'' solutions to the everyday life of users. 
For example, the traditional thermostat, a solution for controlling a room's temperature, can be replaced by a smart thermostat that can extract the users' preferences and can be controlled remotely.

Despite the popularity of IoT solutions, the development of such systems still seems to be a form of art, and the potential issues facing users are largely unknown. 
A systematic identification of problems would enable researchers to devise tools, techniques, and frameworks to support effective development of such systems.
In this paper, we use the user reviews left on the Amazon and Google Play marketplaces to elicit the issues in IoT systems. 
We particularly focus on IoT consumer electronics that are used by home users. 
Most consumer electronics have two main components: a physical device, and a mobile app.
Marketplaces such as Amazon.com and app stores allow users to leave reviews about devices and the mobile apps.

In this paper, we analyze over 90,000 reviews from ten IoT consumer electronic systems to understand the common issues that users are facing. 
We evaluate all reviews from January to mid-October 2018 for ten popular devices from Amazon.com as well as reviews from the corresponding Android apps from Google Play.
Our results indicate that issues with \emph{connectivity}, \emph{timing}, and \emph{updates} are particularly prevalent in the reviews.
The results call for a new software-hardware development framework to assist development of reliable IoT systems.

\noindent\textbf{Contributions.} This paper makes the following contributions.
\begin{itemize}
	\item We identify technical issues in ten consumer IoT systems by analyzing users' reviews on Amazon and Google Play. 
 \item We make data and analysis code available. 
\end{itemize}

\Comment{
\Fix{This can go to the Intro}
Atrozi et al. \cite{Atzori2010}
 survey the definitions, architecture, fundamental technologies, and applications of the Internet of Things.
 They note that IoT has been deployed in the area of mobile apps and that mobile devices will expand the IoT market as they continue to develop, but doesn’t look at mobile IoT apps in much detail; this paper is from 2014, so the research at the time may not have been at this point yet
Points out that IoT’s development faces challenges, such as networks becoming too complex, lack of standards making devices incompatible with one another, and security/privacy protection.

\Fix{It can go to Intro to motivate the study}
2015 – App Store Mining and Analysis (Keynote)
App stores provide new opportunities in scientific research because of how they differ from other software deployment mechanisms
App store reviews provide data on user perceptions, popularity, pricing
In more traditional deployment systems, much of this useful data is usually not commercially released, yet it is publicly available on app stores.
}

%% file: related.tex
\section{Related Work}
There is a large body of work in analyzing users' reviews to elicit the issues in software systems.
To the best of our knowledge, extracting users' issues in IoT technology, at least in the form of consumer electronics, has not been explored. 

Atrozi et al. \cite{Atzori2010} survey the definitions, architecture, fundamental technologies, and applications of the Internet of Things.
 They note that IoT has been deployed in the area of mobile apps and that mobile devices will expand the IoT market as they continue to develop.
 Alur et al.~\cite{alur2016systems} provide a list of challenges in development of IoT systems.
 Fu et al.~\cite{fu2017safety} report the potential safety and security issues in IoT systems.

Maalej and Nabil \cite{Maalej2015} used the probabilistic technique to automatically classify app reviews into one of four types: bug reports, feature requests, user experience and ratings. 
Hoon et al. \cite{Hoon2013} studied how reviews evolve over time and the characteristics of the reviews. 
Pagano and Maalej \cite{Pagano} explore how and when users leave their feedback and also analyzed the content of the review.
AppEcho  allows users to leave feedback in situ that is exactly when the user discovered an issue~\cite{Seyff2014}.
AR-Miner extracts informative information from the reviews using topic modeling~\cite{Chen2014}.

Hermanson~\cite{Hermanson2014} looked at whether perceived ease of use and perceived usefulness were widely discernable in user reviews submitted to apps on the Google Play Store. The author collected 13,099 reviews from the Google Play Store and found that only 3\% of the reviews had information relating to perceived usefulness and that less than 1\% of the reviews had any mention of perceived ease of use.
CLAP is a tool to help developers parse through app reviews when rolling out an update~\cite{Villarroel}.
CLAP categorizes user reviews based on the information they contain.
It groups related reviews together, and then automatically prioritize which groups should be prioritized for the next app update.
Gu and Kim \cite{gu2015parts} propose SUR-Miner, a pattern –based parsing technique which parses aspect-opinion pairs from review sentences to produce effective user review summarization.
Di Sorbo et al.~\cite{DiSorbo2016} propose a tool called SURF which summarizes thousands of app reviews and generates a detailed interactive agenda on recommended updates and changes to the app.
Licorish et al. \cite{licorish2017attributes} used content analysis and regression to provide insights into the nature of reviews provided by the users.
Mujahid et al.~\cite{Mujahid2017} looked at user reviews of wearable apps. The authors manually sampled and categorized six android wearable apps. They found that the most frequent complaints involved functional errors, lack of functionality, and cost.

%% file: method.tex
\section{Method}
\label{sec:method}
In this section, 
we describe the data selection and characteristics of the review data used in this study. 

\subsection{Characteristics of Data}
Table~\ref{tab:appinfo} lists the IoT systems (devices and their corresponding apps) used in this study. 
These systems encompass a wide domain including conversational assistants, thermostats, electronic locks, and tracking devices. 
The price of the devices ranged from about \$25 to \$200 at the time of writing. 
 Six of these systems were used in a previous study of IoT apps by Kaaz et al. ~\cite{kaaz2017understanding}, and  
the remaining four systems are based on a Google search for popular IoT apps. 
For each system, we found an app on Google Play and the corresponding device on \url{Amazon.com}. 
We note that for some of the devices there were multiple versions of the products on Amazon website. 
In such cases, we chose the ones which had more reviews.

\input{produtc-tbl}

For each system, we extracted the reviews from the Amazon website and the corresponding app reviews from the Google Play Store. 
We collected reviews that were posted during a ten-month period starting from the beginning of January 2018 to mid-October of the same year. 

\begin{table}
  \caption{Characteristics of Reviews Considered in this Study}
  \label{tbl:stats}
  \resizebox{\columnwidth}{!}{%
  \begin{tabular}{|*{8}{c|}} \hline
    \multicolumn{2}{c}{System}                     & Total     & \multicolumn{5}{c}{Review Length  (char)}     \\ \hline
    \multicolumn{2}{c}{}                           &           & Min.  & 25\% & 50\% &  75\% & Max \\ \hline
    \multirow{2}{*}{Amazon Alexa}  & App Reviews   & 5,785      & 1 & 18 & 56 & 135 & 2,027  \\ 
                                   & Device Reviews & 54,289         & 3 & 44 & 92 & 192 & 7,632 \\ \hline
\multirow{2}{*}{ecobee}  & App Reviews   & 917      & 4 & 68 & 133 & 229 & 1,572  \\ 
                                   & Device Reviews & 598          & 14 & 148.8 & 336.5 & 644 & 12,390 \\ \hline
\multirow{2}{*}{Google Home}  & App Reviews   & 7,051      & 2 & 26 & 73 & 157 & 1,996  \\ 
                                   & Device Reviews & 1,859          & 9 & 102 & 240 & 468 & 9,526 \\ \hline
\multirow{2}{*}{Insteon}  & App Reviews   & 70      & 7 & 71.5 & 118.5 & 264.8 & 532  \\ 
                                   & Device Reviews &     121      & 19 & 113 & 316 & 621 & 2,232 \\ \hline
\multirow{2}{*}{Kevo}  & App Reviews   &   461    & 3 & 33 & 93 & 206 & 1,724  \\ 
                                   & Device Reviews &     296      & 15 & 154.8 & 337 & 719.2 & 5,016 \\ \hline
\multirow{2}{*}{Nest}  & App Reviews   &  1,798     & 3 & 61 & 135 & 242 & 1,877  \\ 
                                   & Device Reviews &    1,431       & 9 & 83.5 & 210 & 462 & 5,139 \\ \hline
\multirow{2}{*}{Philips Hue}  & App Reviews   &  1,231     & 3 & 64 & 137 & 248 & 1,553  \\ 
                                   & Device Reviews & 667          & 9 & 69 & 146 & 303.5 & 4,833 \\ \hline
\multirow{2}{*}{SmartThings}  & App Reviews   &  9,973     & 2  & 18 & 58 & 139 & 2,662  \\ 
                                   & Device Reviews &   417        & 7 & 89 & 214 & 487 & 3,998 \\ \hline
\multirow{2}{*}{Tile}  & App Reviews   &   1,480    & 2 & 34 & 90.5 & 194 & 1,718  \\ 
                                   & Device Reviews &    2,149       & 7 & 62 & 137 & 256 & 3,209 \\ \hline
\multirow{2}{*}{WeMo}  & App Reviews   &  3,177     & 2 & 40 & 85 & 177 & 1,833  \\ 
                                   & Device Reviews &     2,013      & 5 & 100 & 215 & 385 & 7,841 \\ \hline
  \end{tabular}%
  }
\end{table}

Table~\ref{tbl:stats} shows statistics about the number and length of reviews for the devices and apps.
The table provides some noteworthy insights. For instance, with all IoT systems, the maximum review length was always higher in the device reviews than in the app reviews. It is possible that Amazon allows a higher character limit in its reviews than the Google Play Store. Moreover, users have to use a mobile phone to enter the app reviews, but they can use computers for leaving reviews for the devices on Amazon. It is also possible that typing on a computer can be easier for many users than typing on a phone, leading to longer reviews.

For seven out of ten systems, more reviews were collected from the Google Play Store than Amazon. The three exceptions to this pattern are Amazon Alexa, Insteon, and Tile. With Amazon Alexa, this could be explained by the fact that Amazon is both the creator of the device and the curator of the storefront. As a first-party product, the Echo Dot likely receives some level of favoritism, likely expressed through increased promotion on the Amazon.com web site. This promotion could lead to more purchases and ultimately, more reviews. This favoritism may also explain why the Google Home app received so many more reviews than the Google Home device. The reason Insteon is an exception is probably due to the fact that it received fewer reviews overall. There is only a difference of 51 reviews between the app reviews and the device reviews. If Insteon had received more reviews during the time frame studied, the number of reviews may have more closely matched the pattern of the other systems. With Tile, no explanation for its anomalous behavior is immediately apparent. It is worth noting that Tile, as an IoT system, is fairly unique out of all the systems studied. Tile's functionality is focused on a narrow and specific purpose that none of the other nine systems appear to provide.

\subsection{Topic Modeling}

We used Latent Dirichlet Allocation (LDA) to identify the most important topics users feel most strongly about \cite{fujino2014refining}. By creating topics from the text of these reviews, it is possible that some topics will be comprised of words that speak to a component of the app or device that users are complaining about. For example, if a topic contains the words ``bad'', ``battery'', and ``drain'', then we could infer that complaints about battery life are a significant topic in the user reviews.
We used the Gensim library \cite{gensimPg} with the default configurations to generate a list of topics. 
For each set of reviews, we used LDA to generate three topics and return the ten words for each topic that contributed the most to that topic.

%% file: produtc-tbl.tex
\begin{table}
\caption{IoT Devices and Applications Used in this Study}
\centering
\label{tab:appinfo}
\resizebox{\columnwidth}{!}{%
\begin{tabular}[t]{|l|p{10cm}|} \hline
\makecell{First Row: Name of App \\ \\ Second Row: Name of Device} &   Description
\\ \hline \hline
\makecell{Amazon Alexa \\ \\ Amazon Echo Dot (2nd Gen)} & A virtual assistant. App connects to a variety of devices with speakers and microphones that allows the user to interface with the service.   \\ \hline
\makecell{ecobee \\  \\ ecobee4 Smart Thermostat} & Connects to a thermostat that can be controlled by the app. \\ \hline
\makecell{Google Home \\  \\ Google WiFi System, 1-Pack} & A virtual assistant. App connects to a variety of devices with speakers and microphones that allows the user to interface with the service. \\ \hline
\makecell{Insteon for Hub \\ \\ Insteon Hub} & Connects to a hub device that, in turn, connects to a number of other Insteon devices, including light switches, lamps, and security camera. Through the hub, the user can control all connected devices with the app. \\ \hline
\makecell{Kevo \\ \\ Kevo Lock (2nd Gen)} &  Connects to a door lock that can be installed in the user's door. Lock can be controlled with the app.   \\ \hline
\makecell{Nest \\ \\ Nest T3007ES Thermostat} & Connects to a thermostat that can be controlled by the app. \\ \hline
\makecell{Philips Hue \\ \\ Philips Hue Starter Kit} & Connects to light bulbs whose intensity and color are controlled by the app. \\ \hline
\makecell{SmartThings (Samsung Connect) \\ \\ SmartThingsSmart Home Hub} & Connects to a variety of Samsung-branded devices. These devices can be controlled through the app. \\ \hline
\makecell{Tile \\ \\ Tile Mate} & Connects to a small, square-shaped device that can be attached to a number of personal belongings. The device connects to the internet, allowing its location to be tracked through the app. \\ \hline
\makecell{WeMo \\ \\ WeMo Mini Smart Plug} & Connects to a number of WeMo-branded devices, including cameras, light bulbs, and electrical plugs. These devices can be controlled through the app. \\ \hline
\end{tabular}%
}
\end{table}

%

%% file: issues.tex
\section{Issues mentioned  in IoT System Reviews}
This section describes the result of our analysis of users' reviews for the systems in our study. 
For each IoT system, we generated three topics made up of ten words. Our results listed these ten words in the order of how much they contributed to that topic. For brevity, we discuss the analysis of two systems in detail here. We add the results of the topics discussed in the reviews of other systems in Appendix~\ref{appendix}.

Tables \ref{tab:alexaAPPLDA} and \ref{tab:smartthingsAPPLDA} depict the words for each topic for the Amazon Alexa and SmartThings apps. Tables \ref{tab:alexaHWLDA} and \ref{tab:smartthingsHWLDA} display the topics for the corresponding devices. Beside each word is a number from 0 to 1 that reflects the magnitude at which that word contributed to the topic. When it comes to interpreting the LDA results, it was clear some words in a list appeared to be more important than others. Determining the usefulness of a word was based on a combination of its position in the list and the magnitude value the word had been assigned. A higher magnitude means a word contributed to the topic more strongly, meaning it is likely to be more integral in identifying the topic created by the LDA. At the same time, each topic list spans a different range of values between the magnitude of the first word and the magnitude of the tenth word. In some cases, the final few words had magnitudes so low to appear almost negligible, but in other cases, the final words carried magnitudes not all that lower than the value for the first word in that list.

For example, in Table \ref{tab:alexaHWLDA}, the tenth word in Topic 1 is ``christmas'', which has a magnitude of 0.013. Though  its position near the end of the list means this word may be one of the least important words in Topic 1, its impact is not entirely negligible. Compare the magnitude value of ``christmas'' in Topic 1 to the magnitudes found in Topic 3. The only word in Topic 1 with a magnitude higher than 0.013 is the first word, ``time'', which has a magnitude value of 0.015. Every word following has a lower magnitude value than ``christmas''. This arguably means that ``christmas'' had more of an impact on its topic than nine of the ten words listed for Topic 3. This would suggest that the magnitude values of each word relative to the other magnitude values in the same topic carry more importance than the absolute position in any list.

If an IoT system is receiving significantly different rating distributions from the app store page and device store page, perhaps the kinds of topics generated from the app reviews and the device reviews may illustrate why.

\subsection{Apps vs. Devices}
In a very general sense, the topics for the apps had more instances of words with negative sentiment than the topics for the devices. Though there are plenty of positive words in both the app and device topics, when a negative word like ``slow'', ``bad'', ``waste'', or ``useless'' does appear, it seems to be more likely to be in an app review topic. Additionally, words such as ``control'' and ``connect'' appear more prominently in the app review topics, which may be an indicator of what issues users are running into when using the app. The word ``update'' is particularly common in the app review topics.

\observation{Topics for the apps had more instances of words with negative sentiment than the topics for the devices.}

As an example, none of the topics for the SmartThings Hub device contain any significantly negative language (Table \ref{tab:smartthingsHWLDA}). Meanwhile, the topics for the SmartThings app (Table \ref{tab:smartthingsAPPLDA}) contain significantly more negative language, particularly in Topic 2, where words like ``uninstall'', ``bloatware'', ``remove'', and ``delete'' are all found. The presence of the words ``permission'' and ``update'' in this topic suggest that something about the SmartThings app's permission requirements and updates is being associated with users wanting to remove the app from their device.

Overall, the observations that can be made from these LDA results are fairly general. There are exceptions to the general observations identified above; some negative words do appear in topics for the device reviews, for example. Though the topics provide some guidance as to what kinds of issues users of the apps are facing, it may be possible to refine the results to make these issues more apparent. We decided to see if running an LDA specifically on the app reviews that came with a low star rating might provide more helpful information.
    
\begin{table}
\caption{Amazon Alexa App LDA Topics}
\centering
\resizebox{\columnwidth}{!}{%
\label{tab:alexaAPPLDA}
\begin{tabular}{|c|m{3em}|c|m{3em}|c|m{3em}|} \hline
	Topic 1 Words &   Topic 1 Magnitude & Topic 2 Words &   Topic 2 Magnitude & Topic 3 Words &   Topic 3 Magnitude \\ \hline \hline
	"good" & 0.037 & "love" & 0.021 & "connect" & 0.018 \\ \hline
	"music" & 0.026 & "device" & 0.018 & "time" & 0.016 \\ \hline
	"play" & 0.019 & "update" & 0.016 & "wifi" & 0.015 \\ \hline
	"great" & 0.015 & "slow" & 0.013 & "phone" & 0.014 \\ \hline
	"nice" & 0.011 & "home" & 0.011 & "keep" & 0.013 \\ \hline
	"amazing" & 0.008 & "list" & 0.010 & "update" & 0.012 \\ \hline
	"control" & 0.008 & "awesome" & 0.007 & "android" & 0.009 \\ \hline
	"song" & 0.007 & "take" & 0.007 & "device" & 0.009 \\ \hline
	"voice" & 0.006 & "please" & 0.007 & "tried" & 0.008 \\ \hline
	"time" & 0.006 & "phone" & 0.007 & "best" & 0.008 \\ \hline
\end{tabular}%
}
\end{table}

\begin{table}
\caption{SmartThings App LDA Topics}
\centering
\resizebox{\columnwidth}{!}{%
\label{tab:smartthingsAPPLDA}
\begin{tabular}{|c|m{3em}|c|m{3em}|c|m{3em}|} \hline
	Topic 1 Words &   Topic 1 Magnitude & Topic 2 Words &   Topic 2 Magnitude & Topic 3 Words &   Topic 3 Magnitude \\ \hline \hline
	"great" & 0.035 & "phone" & 0.036 & "tv" & 0.044 \\ \hline
	"love" & 0.024 & "uninstall" & 0.030 & "connect" & 0.028 \\ \hline
	"smartthings" & 0.023 & "permission" & 0.019 & "good" & 0.026 \\ \hline
	"device" & 0.022 & "update" & 0.015 & "device" & 0.020 \\ \hline
	"easy" & 0.016 & "bloatware" & 0.015 & "phone" & 0.017 \\ \hline
	"home" & 0.014 & "disable" & 0.014 & "smart" & 0.015 \\ \hline
	"smart" & 0.013 & "apps" & 0.014 & "time" & 0.013 \\ \hline
	"classic" & 0.013 & "remove" & 0.012 & "bluetooth" & 0.011 \\ \hline
	"useful" & 0.011 & "device" & 0.011 & "update" & 0.011 \\ \hline
	"awesome" & 0.009 & "delete" & 0.011 & "remote" & 0.009 \\ \hline
	\makecell{Topic 1 Summary: \\ Ease of Use }&  & \makecell{Topic 2 Summary: \\ Desire to Remove \\ App from Device }&  & \makecell{Topic 3 Summary: \\ Connecting Phone \\ with App } & \\ \hline
\end{tabular}%
}
\end{table}

\begin{table}
\caption{Amazon Echo Dot LDA Topics}
\centering
\resizebox{\columnwidth}{!}{%
\label{tab:alexaHWLDA}
\begin{tabular}{|c|m{3em}|c|m{3em}|c|m{3em}|} \hline
	Topic 1 Words &   Topic 1 Magnitude & Topic 2 Words &   Topic 2 Magnitude & Topic 3 Words &   Topic 3 Magnitude \\ \hline \hline
	"star" & 0.163 & "music" & 0.040 & "time" & 0.015 \\ \hline
	"five" & 0.116 & "love" & 0.031 & "device" & 0.011 \\ \hline
	"love" & 0.101 & "great" & 0.030 & "know" & 0.011 \\ \hline
	"great" & 0.049 & "speaker" & 0.020 & "answer" & 0.010 \\ \hline
	"fun" & 0.029 & "play" & 0.019 & "question" & 0.010 \\ \hline
	"gift" & 0.027 & "sound" & 0.018 & "voice" & 0.008 \\ \hline
	"easy" & 0.025 & "room" & 0.013 & "ask" & 0.007 \\ \hline
	"product" & 0.020 & "weather" & 0.012 & "phone" & 0.007 \\ \hline
	"four" & 0.020 & "good" & 0.012 & "say" & 0.007 \\ \hline
	"christmas" & 0.013 & "house" & 0.010 & "thing" & 0.007 \\ \hline
	\makecell{Topic 1 Summary: \\ Good Gift \\ for Family}  &  & \makecell{Topic 2 Summary: \\ Good Sound \\ Quality}&  & \makecell{Topic 3 Summary: \\ Voice Interface} & \\ \hline
\end{tabular}%
}
\end{table}

\begin{table}
\caption{SmartThings Hub LDA Topics}
\centering
\resizebox{\columnwidth}{!}{%
\label{tab:smartthingsHWLDA}
\begin{tabular}{|c|m{3em}|c|m{3em}|c|m{3em}|} \hline
	Topic 1 Words &   Topic 1 Magnitude & Topic 2 Words &   Topic 2 Magnitude & Topic 3 Words &   Topic 3 Magnitude \\ \hline \hline
	"light" & 0.013 & "star" & 0.020 & "device" & 0.022 \\ \hline
	"smartthings" & 0.011 & "device" & 0.017 & "smart" & 0.016 \\ \hline
	"turn" & 0.008 & "great" & 0.015 & "home" & 0.013 \\ \hline
	"product" & 0.007 & "home" & 0.014 & "smartthings" & 0.012 \\ \hline
	"home" & 0.007 & "product" & 0.012 & "time" & 0.009 \\ \hline
	"device" & 0.007 & "five" & 0.011 & "light" & 0.008 \\ \hline
	"thing" & 0.006 & "support" & 0.007 & "lock" & 0.008 \\ \hline
	"good" & 0.006 & "smartthings" & 0.007 & "easy" & 0.007 \\ \hline
	"sensor" & 0.005 & "smart" & 0.006 & "support" & 0.007 \\ \hline
	"lot" & 0.005 & "setup" & 0.006 & "great" & 0.006 \\ \hline
\end{tabular}%
}
\end{table}

\subsection{Issues in low-rated systems}

We filtered the app reviews so only reviews that had a minimal 1-star rating were left in the text. 
The goal behind running the LDA on only the 1-star reviews was to see if it was possible to identify the aspects of the apps and devices that were leaving users with a negative impression. As such, we did not focus on words dealing with sentiment or emotion. Instead, we looked at words related to the functionality and features of the apps and devices. Table \ref{tbl:1starLDA} shows some of the noteworthy words that appeared in the topics for each app. Table \ref{tbl:1starLDAHW} shows the same, but for words from device review topics. These are words that stood out for having relatively high magnitude values or for appearing in multiple topics.

Going over all the topics, a handful of relevant words seemed to appear with a greater frequency than others in the apps. For example, for all apps except for Kevo, at least one topic contained either the word ``connect'' or ``connection''. The prevalence of these words suggests that users of these apps have experienced some issue with connecting their phone to another device or network. The frequency in which ``connect'' and ``connection'' appears can mean that these connection issues are perhaps a greater source of frustration for users of IoT apps in general. Another noteworthy word was ``update''. This word appeared in topics for all apps except for Insteon for Hub and Tile. It is important to note that the context for this word may not be the same in every appearance in the tables. For example, it is possible that some topics use ``update'', because an update was the source of a problem. It is also possible that the word appears in the context of users requesting an update to fix a problem with the app. However, the prevalence of the word does indicate that updates are an important part of app development and care should be taken in determining how they are implemented.

``Home'' was another common word that appeared for six apps. With Google Home, this is not all that surprising, since ``home'' is part of the app's name. As for the other apps, the frequency of the word might suggest that many of these apps are indeed utilized for personal, home use. Making sure that these apps remain suited to this kind of use is another important thing for developers to keep in mind.

The word that appeared with the greatest frequency, however, was ``time''. This word appeared in at least one topic for all ten apps. With the exceptions of SmartThings and Insteon, ``time'' actually appeared in at least two of the three topics for every app. Similar to ``update'', ``time''  does not necessarily have a single meaning in every one of its appearances. For apps like Philips Hue, the word appears to refer to the user's ability to configure through the app the time in which their light bulbs are set to turn on, turn off, change color, and so on. In these cases, the word ``time'' seems to relate more to scheduling functions of the app. In other cases, such as with Amazon Alexa, ``time'' appears in conjunction with words like ``slow''. Here, ``time'' seems to be used to refer more to the duration of a function. The word appears in at least one of these contexts for every app. The  prevalence of the word suggests that issues involving time are also an important element of these low-rated reviews. Resolving issues involved with timing settings as well as working to reduce the duration of app functions appear to both be issues app developers may want to pay attention to.

\observation{Issues with \emph{connectivity}, \emph{timing} and \emph{update} are  prevalent in the reviews of apps. }

In the 1-star device reviews, in addition to mentions of timing and connectivity, the word ``support'' is also prominent, appearing in topics for eight of the ten devices. Again, the word seems to have different meanings based on its context. In some cases, ``support'' appears to be related to customer support concerns. In other cases, the word seems to refer to whether the device is still supprorted by the developer. For example, a user may complain that their device is no longer compatible with the latest version of the app. 


In a fast-paced market such as IoT, abandonment of a product is something that might happen, but it is far from ideal. 
This kind of abandonment might suggest that the initial design of a system does not always account for efficient maintenance of the system.
Unsupported devices, also known as zombie devices pose serious security, privacy and safety threats to the users~\cite{fu2017safety}

\observation{Issues with \emph{connectivity}, \emph{timing}, and \emph{support} are  prevalent in the reviews of the device.}

\begin{table}
  \caption{Prominent Words from LDA Topics of 1-Star App Reviews}
  \label{tbl:1starLDA}
  \resizebox{\columnwidth}{!}{%
  \begin{tabular}{|*{7}{c|}} \hline
    System                   & \multicolumn{5}{c}{Words}     \\ \hline
                                          & 1  & 2 & 3 &  4 & 5 \\ \hline
    Amazon Alexa  & hate     & device & time & update & useless  \\ 
\hline
ecobee     & update      & thermostat & time & internet & connection   \\ 
\hline
Google Home     &  music     & chromecast & time & device & update   \\ 
\hline
Insteon  &    device    & time & find & waste & version   \\ 
\hline
Kevo     &   lock    & update & door & phone & time   \\ 
\hline
Nest     &  camera     & thermostat & update & home & time   \\ 
\hline
Philips Hue     &  light     & update & bridge & time & connection   \\ 
\hline
SmartThings     &  phone     & permission  & uninstall & access & connect   \\ 
\hline
Tile    &   phone    & time & find & battery & key   \\ 
\hline
WeMo     &  device     & time & product & switch & update   \\ 
\hline
  \end{tabular}%
  }
\end{table}

\begin{table}
  \caption{Prominent Words from LDA Topics of 1-Star Device Reviews}
  \label{tbl:1starLDAHW}
  \resizebox{\columnwidth}{!}{%
  \begin{tabular}{|*{7}{c|}} \hline
    System                   & \multicolumn{5}{c}{Words}     \\ \hline
                                          & 1  & 2 & 3 &  4 & 5 \\ \hline
    Amazon Alexa  &  time    & device & star & music & sound  \\ 
\hline
ecobee     &   thermostat    & support & product & system & temperature   \\ 
\hline
Google Home     &    wifi   & device & router & product & support   \\ 
\hline
Insteon  &    support    & device & customer & sensor & year   \\ 
\hline
Kevo     &    lock   & door & phone & time & product   \\ 
\hline
Nest     &    thermostat   & support & product & time & heat   \\ 
\hline
Philips Hue     &    bulb   & light & bridge & support & turn   \\ 
\hline
SmartThings     & product      & device  & home & time &  new  \\ 
\hline
Tile    &    phone   & battery & key & time & product   \\ 
\hline
WeMo     &  device     & switch & connect & smart & time   \\ 
\hline
  \end{tabular}%
  }
\end{table}

%% file: discussion.tex
\section{Discussion}

The intent behind running topic modeling on the app and device reviews was to help identify those functions and features of the IoT system that appeared to be the most important to its users. After seeing the greater distribution of 1-star reviews in the apps compared to the devices, we were interested in discovering whether the LDA results would in particular help identify the characteristics of the apps that were causing users to leave negative reviews. The topics generated by the LDA from each of the review texts provided fairly general information. Negative words appeared to be more common in the app review topics than in the device review topics, for example.

Running LDA on the 1-star app reviews only seemed to produce slightly more tangible results. Words like ``time'', ``update'', and ``connect'' were particularly frequent among these topics. Each of these words is related to different aspects of an app's functionality that can be a focus for developers. Though it is likely that the process can be refined further to be more effective, the results suggest that topic modeling approaches such as LDA can be used to help identify issues users may be dealing with when using an IoT system.

The three prominent issues of timing, connectivity, and updates shed light on some facets of IoT systems that are rarely encountered in developing mainstream software systems.
Powerful processors, abundant memory, and optimizing compilers have largely resolved the problem of timing and efficiency in the development of software. 
However, in systems that work on limited processing power and memory such as IoT devices and the mobile systems, efficiency has become an issue. 

Moreover, fast, reliable networks with negligible latency are a given in the development of traditional software systems.
This has been achieved by development of technologies and tools that reduce the latency of network connections;
for example, nowadays, almost all cloud service providers automatically move the running instances of applications to data centers closer to clients.
It seems that we need new technologies to address this problem for IoT systems. 

The problem of automatic updates and backwards compatibility in traditional software systems have been under investigation for many years.
Nowadays, thanks to standardization of operating systems and protocols, there are frameworks that strive to (almost) achieve seamless updates of software. 
For example, Android, Windows, and MacOS allow developers to update their applications using the corresponding app stores. 
However, updates for IoT systems for which a large portion of the hardware and  protocols have not been standardized pose new challenges that require new tools and techniques. 

Understanding issues and obstacles in operational IoT systems allows us to devise techniques and tools to support effective development of these systems. 
We believe that analysis of user reveiews can contribute to a better understanding of these systems by extracting first-hand experiences of users.
We released the dataset and the source code of this study at \url{https://github.com/atruelove/AppReviewAnalysis} to replicate the study and to facilitate further analysis of the reviews.

%% file: threats.tex
\section{Threats to Validity}

These are the following main threats to the validity of this study.
First, our analysis was small in scope, we only used relatively recent reviews of a small number of  IoT systems in our study.
We also included the reviews from the Google Play app store but not from other app stores. 
Although small in scope, we believe that this study will provide the first glimpse of the users' issues in IoT systems.
Second, we used LDA for topic modeling. It is known that LDA suffers from some limitations such as order effect~\cite{LDAProblem}.  
To address these limitations, for given proposed words as topics, we manually checked the words to understand the intended meaning in the reviews and make sense of them.

%% file: conclusion.tex
\section{Conclusion}


In this paper, we analyzed the reviews of ten IoT devices from Amazon and the reviews of the corresponding apps from the Google Play Store.
To the best of our knowledge, it is the first analysis of such systems. 
Our results suggest that (1) there are more negative topics in the mobile apps than the devices, and (2) efficiency, connectivity, and updates seem to be prevalent issues in such systems. 
Our results call for the development of new tools and techniques to support practitioners to address these issues. 
We released the dataset and the source code of this study at \url{https://github.com/atruelove/AppReviewAnalysis} to facilitate further analysis of the reviews. 

%% file: Acknowledgment.tex
\noindent\textbf{Acknowledgment}
We would like to thank the anonymous reviewers. We would also like to thank Soodeh Atefi,  and Md Rafiqul Islam Rabin for their  comments on the earlier versions of this paper.

%% file: extra.tex
\begin{table}
\caption{ecobee App}
\centering
\resizebox{\columnwidth}{!}{%
\label{tab:ecobeeAPPLDA}
\begin{tabular}{|c|m{3em}|c|m{3em}|c|m{3em}|} \hline
	Topic 1 Words &   Topic 1 Magnitude & Topic 2 Words &   Topic 2 Magnitude & Topic 3 Words &   Topic 3 Magnitude \\ \hline \hline
	"update" & 0.033 & "great" & 0.015 & "update" & 0.024 \\ \hline
	"thermostat" & 0.023 & "love" & 0.013 & "thermostat" & 0.022 \\ \hline
	"new" & 0.014 & "back" & 0.012 & "time" & 0.016 \\ \hline
	"version" & 0.012 & "thermostat" & 0.011 & "connection" & 0.015 \\ \hline
	"great" & 0.012 & "geofencing" & 0.011 & "internet" & 0.015 \\ \hline
	"sensor" & 0.011 & "geofence" & 0.010 & "setting" & 0.010 \\ \hline
	"tablet" & 0.010 & "update" & 0.009 & "log" & 0.009 \\ \hline
	"back" & 0.009 & "much" & 0.009 & "screen" & 0.009 \\ \hline
	"phone" & 0.009 & "time" & 0.008 & "device" & 0.009 \\ \hline
	"screen" & 0.009 & "setting" & 0.008 & "wifi" & 0.009 \\ \hline
\end{tabular}%
}
\end{table}

\begin{table}
\caption{Google Home App}
\centering
\resizebox{\columnwidth}{!}{%
\label{tab:googleAPPLDA}
\begin{tabular}{|c|m{3em}|c|m{3em}|c|m{3em}|} \hline
	Topic 1 Words &   Topic 1 Magnitude & Topic 2 Words &   Topic 2 Magnitude & Topic 3 Words &   Topic 3 Magnitude \\ \hline \hline
	"good" & 0.067 & "home" & 0.039 & "chromecast" & 0.039 \\ \hline
	"great" & 0.050 & "love" & 0.015 & "cast" & 0.029 \\ \hline
	"nice" & 0.019 & "device" & 0.014 & "device" & 0.026 \\ \hline
	"awesome" & 0.018 & "mini" & 0.011 & "update" & 0.022 \\ \hline
	"easy" & 0.013 & "music" & 0.010 & "phone" & 0.017 \\ \hline
	"ok" & 0.009 & "control" & 0.010 & "tv" & 0.016 \\ \hline
	"video" & 0.006 & "play" & 0.009 & "connect" & 0.015 \\ \hline
	"excellent" & 0.006 & "voice" & 0.008 & "wifi" & 0.012 \\ \hline
	"thank" & 0.005 & "speaker" & 0.007 & "screen" & 0.012 \\ \hline
	"thanks" & 0.005 & "add" & 0.007 & "time" & 0.012 \\ \hline
\end{tabular}%
}
\end{table}

\begin{table}
\caption{Insteon App}
\centering
\resizebox{\columnwidth}{!}{%
\label{tab:insteonAPPLDA}
\begin{tabular}{|c|m{3em}|c|m{3em}|c|m{3em}|} \hline
	Topic 1 Words &   Topic 1 Magnitude & Topic 2 Words &   Topic 2 Magnitude & Topic 3 Words &   Topic 3 Magnitude \\ \hline \hline
	"device" & 0.018 & "thermostat" & 0.008 & "support" & 0.010 \\ \hline
	"new" & 0.014 & "device" & 0.008 & "interface" & 0.010 \\ \hline
	"time" & 0.012 & "keep" & 0.006 & "hardware" & 0.008 \\ \hline
	"back" & 0.010 & "time" & 0.006 & "account" & 0.008 \\ \hline
	"connect" & 0.010 & "notification" & 0.006 & "product" & 0.008 \\ \hline
	"money" & 0.008 & "setting" & 0.006 & "update" & 0.008 \\ \hline
	"slow" & 0.008 & "never" & 0.006 & "let" & 0.008 \\ \hline
	"waste" & 0.008 & "useless" & 0.006 & "oh" & 0.008 \\ \hline
	"good" & 0.008 & "constantly" & 0.006 & "month" & 0.008 \\ \hline
	"horrible" & 0.007 & "multiple" & 0.006 & "device" & 0.008 \\ \hline
\end{tabular}%
}
\end{table}

\begin{table}
\caption{Kevo App}
\centering
\resizebox{\columnwidth}{!}{%
\label{tab:kevoAPPLDA}
\begin{tabular}{|c|m{3em}|c|m{3em}|c|m{3em}|} \hline
	Topic 1 Words &   Topic 1 Magnitude & Topic 2 Words &   Topic 2 Magnitude & Topic 3 Words &   Topic 3 Magnitude \\ \hline \hline
	"lock" & 0.046 & "lock" & 0.032 & "lock" & 0.017 \\ \hline
	"door" & 0.024 & "phone" & 0.024 & "time" & 0.016 \\ \hline
	"time" & 0.022 & "android" & 0.017 & "great" & 0.016 \\ \hline
	"phone" & 0.018 & "great" & 0.015 & "good" & 0.016 \\ \hline
	"update" & 0.017 & "open" & 0.012 & "easy" & 0.010 \\ \hline
	"key" & 0.014 & "update" & 0.012 & "pretty" & 0.009 \\ \hline
	"unlock" & 0.011 & "time" & 0.008 & "battery" & 0.009 \\ \hline
	"love" & 0.009 & "note" & 0.007 & "key" & 0.009 \\ \hline
	"open" & 0.008 & "device" & 0.007 & "awesome" & 0.008 \\ \hline
	"connect" & 0.008 & "running" & 0.007 & "update" & 0.006 \\ \hline
\end{tabular}%
}
\end{table}

\begin{table}
\caption{Nest App}
\centering
\resizebox{\columnwidth}{!}{%
\label{tab:nestAPPLDA}
\begin{tabular}{|c|m{3em}|c|m{3em}|c|m{3em}|} \hline
	Topic 1 Words &   Topic 1 Magnitude & Topic 2 Words &   Topic 2 Magnitude & Topic 3 Words &   Topic 3 Magnitude \\ \hline \hline
	"camera" & 0.020 & "home" & 0.033 & "update" & 0.025 \\ \hline
	"time" & 0.016 & "away" & 0.022 & "camera" & 0.021 \\ \hline
	"thermostat" & 0.016 & "great" & 0.014 & "load" & 0.014 \\ \hline
	"notification" & 0.015 & "turn" & 0.014 & "video" & 0.012 \\ \hline
	"update" & 0.010 & "thermostat" & 0.014 & "connect" & 0.011 \\ \hline
	"offline" & 0.008 & "phone" & 0.011 & "take" & 0.010 \\ \hline
	"issue" & 0.006 & "love" & 0.010 & "wifi" & 0.010 \\ \hline
	"phone" & 0.005 & "time" & 0.009 & "last" & 0.009 \\ \hline
	"android" & 0.005 & "temperature" & 0.009 & "fix" & 0.008 \\ \hline
	"product" & 0.005 & "product" & 0.008 & "great" & 0.008 \\ \hline
\end{tabular}%
}
\end{table}

\begin{table}
\caption{Philips Hue App}
\centering
\resizebox{\columnwidth}{!}{%
\label{tab:philipshueAPPLDA}
\begin{tabular}{|c|m{3em}|c|m{3em}|c|m{3em}|} \hline
	Topic 1 Words &   Topic 1 Magnitude & Topic 2 Words &   Topic 2 Magnitude & Topic 3 Words &   Topic 3 Magnitude \\ \hline \hline
	"light" & 0.041 & "version" & 0.011 & "update" & 0.031 \\ \hline
	"room" & 0.016 & "light" & 0.010 & "bridge" & 0.028 \\ \hline
	"great" & 0.014 & "color" & 0.007 & "light" & 0.023 \\ \hline
	"update" & 0.013 & "location" & 0.007 & "home" & 0.018 \\ \hline
	"scene" & 0.012 & "scene" & 0.007 & "connect" & 0.014 \\ \hline
	"time" & 0.011 & "routine" & 0.007 & "time" & 0.008 \\ \hline
	"bulb" & 0.010 & "good" & 0.007 & "control" & 0.008 \\ \hline
	"new" & 0.010 & "gen" & 0.006 & "connection" & 0.008 \\ \hline
	"turn" & 0.010 & "easy" & 0.006 & "new" & 0.007 \\ \hline
	"feature" & 0.009 & "feature" & 0.006 & "find" & 0.006 \\ \hline
\end{tabular}%
}
\end{table}

\begin{table}
\caption{Tile App}
\centering
\resizebox{\columnwidth}{!}{%
\label{tab:tileAPPLDA}
\begin{tabular}{|c|m{3em}|c|m{3em}|c|m{3em}|} \hline
	Topic 1 Words &   Topic 1 Magnitude & Topic 2 Words &   Topic 2 Magnitude & Topic 3 Words &   Topic 3 Magnitude \\ \hline \hline
	"great" & 0.021 & "love" & 0.020 & "phone" & 0.039 \\ \hline
	"tile" & 0.018 & "phone" & 0.018 & "key" & 0.029 \\ \hline
	"find" & 0.016 & "found" & 0.010 & "find" & 0.020 \\ \hline
	"battery" & 0.016 & "key" & 0.009 & "time" & 0.017 \\ \hline
	"time" & 0.014 & "easy" & 0.008 & "tile" & 0.013 \\ \hline
	"phone" & 0.014 & "find" & 0.008 & "keep" & 0.010 \\ \hline
	"key" & 0.011 & "location" & 0.008 & "lost" & 0.009 \\ \hline
	"product" & 0.010 & "great" & 0.008 & "location" & 0.008 \\ \hline
	"never" & 0.009 & "best" & 0.008 & "bluetooth" & 0.008 \\ \hline
	"year" & 0.008 & "good" & 0.007 & "ring" & 0.008 \\ \hline
\end{tabular}%
}
\end{table}

\begin{table}
\caption{WeMo App}
\centering
\resizebox{\columnwidth}{!}{%
\label{tab:wemoAPPLDA}
\begin{tabular}{|c|m{3em}|c|m{3em}|c|m{3em}|} \hline
	Topic 1 Words &   Topic 1 Magnitude & Topic 2 Words &   Topic 2 Magnitude & Topic 3 Words &   Topic 3 Magnitude \\ \hline \hline
	"switch" & 0.025 & "great" & 0.037 & "device" & 0.035 \\ \hline
	"light" & 0.024 & "easy" & 0.033 & "time" & 0.023 \\ \hline
	"update" & 0.021 & "time" & 0.027 & "wifi" & 0.019 \\ \hline
	"home" & 0.021 & "device" & 0.018 & "connect" & 0.019 \\ \hline
	"turn" & 0.017 & "setup" & 0.017 & "switch" & 0.013 \\ \hline
	"device" & 0.015 & "good" & 0.015 & "setup" & 0.012 \\ \hline
	"smart" & 0.012 & "love" & 0.013 & "rule" & 0.011 \\ \hline
	"plug" & 0.011 & "buggy" & 0.012 & "network" & 0.011 \\ \hline
	"firmware" & 0.011 & "slow" & 0.009 & "plug" & 0.011 \\ \hline
	"product" & 0.011 & "product" & 0.008 & "reset" & 0.009 \\ \hline
\end{tabular}%
}
\end{table}

\begin{table}
\caption{ecobee4 Smart Thermostat}
\centering
\resizebox{\columnwidth}{!}{%
\label{tab:ecobeeHWLDA}
\begin{tabular}{|c|m{3em}|c|m{3em}|c|m{3em}|} \hline
	Topic 1 Words &   Topic 1 Magnitude & Topic 2 Words &   Topic 2 Magnitude & Topic 3 Words &   Topic 3 Magnitude \\ \hline \hline
	"support" & 0.013 & "thermostat" & 0.021 & "thermostat" & 0.030 \\ \hline
	"thermostat" & 0.012 & "great" & 0.016 & "easy" & 0.014 \\ \hline
	"unit" & 0.008 & "sensor" & 0.012 & "great" & 0.012 \\ \hline
	"system" & 0.007 & "love" & 0.012 & "sensor" & 0.011 \\ \hline
	"time" & 0.007 & "product" & 0.011 & "install" & 0.010 \\ \hline
	"sensor" & 0.006 & "room" & 0.011 & "smart" & 0.009 \\ \hline
	"product" & 0.006 & "home" & 0.010 & "house" & 0.008 \\ \hline
	"customer" & 0.006 & "easy" & 0.009 & "device" & 0.008 \\ \hline
	"new" & 0.005 & "smart" & 0.008 & "wire" & 0.008 \\ \hline
	"back" & 0.005 & "temperature" & 0.008 & "home" & 0.007 \\ \hline
\end{tabular}%
}
\end{table}

\begin{table}
\caption{Google WiFi System}
\centering
\resizebox{\columnwidth}{!}{%
\label{tab:googleHWLDA}
\begin{tabular}{|c|m{3em}|c|m{3em}|c|m{3em}|} \hline
	Topic 1 Words &   Topic 1 Magnitude & Topic 2 Words &   Topic 2 Magnitude & Topic 3 Words &   Topic 3 Magnitude \\ \hline \hline
	"easy" & 0.045 & "device" & 0.025 & "wifi" & 0.034 \\ \hline
	"great" & 0.039 & "network" & 0.018 & "router" & 0.024 \\ \hline
	"wifi" & 0.034 & "wifi" & 0.013 & "speed" & 0.013 \\ \hline
	"house" & 0.030 & "mesh" & 0.009 & "device" & 0.011 \\ \hline
	"star" & 0.022 & "router" & 0.008 & "house" & 0.010 \\ \hline
	"setup" & 0.019 & "point" & 0.007 & "internet" & 0.009 \\ \hline
	"five" & 0.017 & "phone" & 0.007 & "system" & 0.009 \\ \hline
	"coverage" & 0.016 & "connected" & 0.006 & "network" & 0.008 \\ \hline
	"signal" & 0.014 & "setup" & 0.006 & "setup" & 0.008 \\ \hline
	"love" & 0.013 & "house" & 0.006 & "home" & 0.008 \\ \hline
\end{tabular}%
}
\end{table}

\begin{table}
\caption{Insteon Hub Device}
\centering
\resizebox{\columnwidth}{!}{%
\label{tab:insteonHWLDA}
\begin{tabular}{|c|m{3em}|c|m{3em}|c|m{3em}|} \hline
	Topic 1 Words &   Topic 1 Magnitude & Topic 2 Words &   Topic 2 Magnitude & Topic 3 Words &   Topic 3 Magnitude \\ \hline \hline
	"device" & 0.020 & "new" & 0.016 & "device" & 0.013 \\ \hline
	"year" & 0.017 & "device" & 0.015 & "product" & 0.013 \\ \hline
	"switch" & 0.011 & "great" & 0.012 & "support" & 0.008 \\ \hline
	"product" & 0.011 & "scene" & 0.012 & "new" & 0.008 \\ \hline
	"new" & 0.008 & "sensor" & 0.011 & "home" & 0.007 \\ \hline
	"control" & 0.008 & "light" & 0.009 & "star" & 0.006 \\ \hline
	"account" & 0.008 & "switch" & 0.007 & "customer" & 0.006 \\ \hline
	"light" & 0.007 & "support" & 0.007 & "control" & 0.006 \\ \hline
	"support" & 0.007 & "problem" & 0.006 & "de" & 0.005 \\ \hline
	"time" & 0.006 & "home" & 0.006 & "wish" & 0.005 \\ \hline
\end{tabular}%
}
\end{table}

\begin{table}
\caption{Kevo Lock (2nd Gen)}
\centering
\resizebox{\columnwidth}{!}{%
\label{tab:kevoHWLDA}
\begin{tabular}{|c|m{3em}|c|m{3em}|c|m{3em}|} \hline
	Topic 1 Words &   Topic 1 Magnitude & Topic 2 Words &   Topic 2 Magnitude & Topic 3 Words &   Topic 3 Magnitude \\ \hline \hline
	"lock" & 0.046 & "lock" & 0.020 & "lock" & 0.046 \\ \hline
	"door" & 0.014 & "time" & 0.015 & "door" & 0.017 \\ \hline
	"time" & 0.013 & "product" & 0.014 & "key" & 0.015 \\ \hline
	"key" & 0.013 & "phone" & 0.012 & "phone" & 0.013 \\ \hline
	"phone" & 0.010 & "door" & 0.010 & "time" & 0.010 \\ \hline
	"battery" & 0.010 & "great" & 0.008 & "touch" & 0.008 \\ \hline
	"open" & 0.008 & "star" & 0.006 & "unlock" & 0.007 \\ \hline
	"great" & 0.007 & "support" & 0.006 & "open" & 0.007 \\ \hline
	"unlock" & 0.006 & "back" & 0.006 & "battery" & 0.006 \\ \hline
	"easy" & 0.006 & "bluetooth" & 0.006 & "product" & 0.006 \\ \hline
\end{tabular}%
}
\end{table}

\begin{table}
\caption{Nest T3007ES Thermostat}
\centering
\resizebox{\columnwidth}{!}{%
\label{tab:nestHWLDA}
\begin{tabular}{|c|m{3em}|c|m{3em}|c|m{3em}|} \hline
	Topic 1 Words &   Topic 1 Magnitude & Topic 2 Words &   Topic 2 Magnitude & Topic 3 Words &   Topic 3 Magnitude \\ \hline \hline
	"thermostat" & 0.031 & "star" & 0.034 & "thermostat" & 0.022 \\ \hline
	"easy" & 0.028 & "great" & 0.029 & "support" & 0.012 \\ \hline
	"install" & 0.017 & "five" & 0.025 & "wire" & 0.012 \\ \hline
	"love" & 0.017 & "product" & 0.020 & "product" & 0.009 \\ \hline
	"home" & 0.015 & "thermostat" & 0.013 & "unit" & 0.009 \\ \hline
	"temperature" & 0.011 & "home" & 0.011 & "system" & 0.008 \\ \hline
	"great" & 0.011 & "temperature" & 0.009 & "time" & 0.008 \\ \hline
	"house" & 0.010 & "love" & 0.006 & "issue" & 0.007 \\ \hline
	"control" & 0.007 & "good" & 0.006 & "customer" & 0.007 \\ \hline
	"money" & 0.007 & "degree" & 0.006 & "service" & 0.007 \\ \hline
\end{tabular}%
}
\end{table}

\begin{table}
\caption{Philips Hue Starter Kit}
\centering
\resizebox{\columnwidth}{!}{%
\label{tab:philipshueHWLDA}
\begin{tabular}{|c|m{3em}|c|m{3em}|c|m{3em}|} \hline
	Topic 1 Words &   Topic 1 Magnitude & Topic 2 Words &   Topic 2 Magnitude & Topic 3 Words &   Topic 3 Magnitude \\ \hline \hline
	"star" & 0.050 & "light" & 0.046 & "light" & 0.037 \\ \hline
	"five" & 0.035 & "great" & 0.034 & "bulb" & 0.022 \\ \hline
	"product" & 0.021 & "bulb" & 0.030 & "turn" & 0.015 \\ \hline
	"bulb" & 0.018 & "home" & 0.027 & "love" & 0.009 \\ \hline
	"love" & 0.011 & "easy" & 0.024 & "bridge" & 0.008 \\ \hline
	"great" & 0.010 & "smart" & 0.015 & "switch" & 0.007 \\ \hline
	"good" & 0.010 & "turn" & 0.014 & "color" & 0.007 \\ \hline
	"easy" & 0.009 & "love" & 0.013 & "router" & 0.007 \\ \hline
	"light" & 0.009 & "setup" & 0.010 & "connect" & 0.006 \\ \hline
	"four" & 0.007 & "control" & 0.009 & "system" & 0.006 \\ \hline
\end{tabular}%
}
\end{table}

\begin{table}
\caption{Tile Mate}
\centering
\resizebox{\columnwidth}{!}{%
\label{tab:tileHWLDA}
\begin{tabular}{|c|m{3em}|c|m{3em}|c|m{3em}|} \hline
	Topic 1 Words &   Topic 1 Magnitude & Topic 2 Words &   Topic 2 Magnitude & Topic 3 Words &   Topic 3 Magnitude \\ \hline \hline
	"battery" & 0.045 & "star" & 0.050 & "phone" & 0.043 \\ \hline
	"year" & 0.028 & "key" & 0.042 & "key" & 0.037 \\ \hline
	"month" & 0.023 & "five" & 0.037 & "find" & 0.031 \\ \hline
	"new" & 0.014 & "great" & 0.027 & "easy" & 0.014 \\ \hline
	"tile" & 0.013 & "time" & 0.023 & "great" & 0.012 \\ \hline
	"good" & 0.012 & "gift" & 0.021 & "love" & 0.012 \\ \hline
	"product" & 0.012 & "love" & 0.017 & "time" & 0.011 \\ \hline
	"stopped" & 0.010 & "phone" & 0.015 & "product" & 0.010 \\ \hline
	"last" & 0.010 & "find" & 0.010 & "bluetooth" & 0.009 \\ \hline
	"buy" & 0.010 & "wallet" & 0.009 & "lose" & 0.007 \\ \hline
\end{tabular}%
}
\end{table}

\begin{table}
\caption{WeMo Mini Smart Plug}
\centering
\resizebox{\columnwidth}{!}{%
\label{tab:wemoHWLDA}
\begin{tabular}{|c|m{3em}|c|m{3em}|c|m{3em}|} \hline
	Topic 1 Words &   Topic 1 Magnitude & Topic 2 Words &   Topic 2 Magnitude & Topic 3 Words &   Topic 3 Magnitude \\ \hline \hline
	"switch" & 0.019 & "connect" & 0.039 & "great" & 0.039 \\ \hline
	"great" & 0.018 & "network" & 0.037 & "time" & 0.029 \\ \hline
	"light" & 0.016 & "router" & 0.028 & "device" & 0.028 \\ \hline
	"wifi" & 0.014 & "support" & 0.017 & "wifi" & 0.028 \\ \hline
	"easy" & 0.014 & "mini" & 0.016 & "internet" & 0.028 \\ \hline
	"turn" & 0.014 & "device" & 0.015 & "pain" & 0.027 \\ \hline
	"plug" & 0.012 & "plug" & 0.015 & "recommend" & 0.022 \\ \hline
	"product" & 0.012 & "service" & 0.014 & "setting" & 0.021 \\ \hline
	"device" & 0.011 & "ssid" & 0.014 & "reset" & 0.021 \\ \hline
	"smart" & 0.010 & "enable" & 0.014 & "easy" & 0.020 \\ \hline
\end{tabular}%
}
\end{table}